# Twist-Free Enhancement of Strength and Modulus in Electrospun Yarns via Liquid-Assisted Capillary Densification

Saujatya Mandal, Sonu Dhiman, Debashish Das[*]

Department of Mechanical Engineering, Indian Institute of Science, Bengaluru, Karnataka 560012, India

## Abstract

Electrospun yarns often fall short of the strength and stiffness of their constituent nanofibers because of loose packing and inter-fiber slip. We report a simple, twist-free route to close this gap by liquid-assisted rolling: yarns are briefly wetted (water or ethanol) and subjected to gentle rolling action (mechanical strokes both perpendicular and parallel to the yarn axis), then dried under controlled conditions so that meniscus forces compact the assembly into tightly bound bundles. The treatment yields large gains in tensile strength and modulus, and as yarn diameter decreases the properties of liquid-treated yarns approach single-fiber limits, indicating more efficient load transfer. Dry-rolling controls produce negligible changes compared to as-spun yarns, confirming that capillarity-driven consolidation—rather than mechanical pressing—dominates the improvement. Water consistently outperforms ethanol, reflecting its larger elastocapillary driving term γ(1+cos θ) on PAN and thus stronger capillary compaction; a short post-treatment anneal near $T_g$ further increases stiffness with a corresponding reduction in ductility. To rationalize these trends, we quantify microstructure via SEM-derived alignment and packing density and show that these complementary descriptors jointly explain variability in mechanical response. A compact constitutive framework, grounded in distributed fiber recruitment and adhesion/frictional contact, captures the observed strengthening–ductility trade-off across processing routes. The results establish capillarity-driven consolidation as a scalable pathway to engineer processing–structure–property relationships in hierarchical polymer fiber assemblies and provide practical guidance for upgrading electrospun yarns—alone or as precursors to twisted and composite architectures.

## 1. Introduction

Electrospinning has emerged as a versatile and scalable technique for fabricating polymer nanofibers with diameters ranging from tens of nanometers to a few microns, offering exceptional surface-to-volume ratios, controllable porosity, and favorable mechanical properties [1–3]. These attributes have positioned electrospun nanofibers as attractive candidates for diverse applications, including filtration membranes, energy storage devices, biomedical scaffolds, and flexible electronics [4–7]. Among electrospinnable polymers, polyacrylonitrile (PAN) stands out for its unique ability to undergo homogeneous deformation under uniaxial tension, effectively avoiding strain localization and necking [8, 9]. In addition to its ease of processing, PAN is widely recognized as a critical precursor for the fabrication of high-performance carbon fibers [10], further underscoring its importance in advanced materials applications.

[*] Corresponding Author: ddas@iisc.ac.in

To exploit the intrinsic mechanical properties of individual electrospun nanofibers in macroscale assemblies, recent research has increasingly focused on assembling fibers into yarns and bundles [1, 11, 12]. Despite these efforts, electrospun yarns typically exhibit mechanical performance far below that of their constituent fibers, primarily due to low packing densities, misalignment, and inefficient interfiber load transfer mechanisms [13, 14]. Traditional approaches to mitigate these limitations, including mechanical twisting [15], post-drawing treatments [16], and solvent fusion techniques [17], have shown improvements in alignment and cohesion but suffer from added complexity, scalability issues, and often inconsistent structural outcomes.

An alternative and promising route to enhancing yarn integrity is capillary-driven densification, in which capillary forces—arising during the evaporation of a liquid [18, 19] from a fiber assembly—actively pull fibers together to promote consolidation and alignment [20–22]. While this mechanism has been exploited to organize carbon nanotube forests and microstructured arrays [20, 21, 23], it has received little attention for electrospun yarns or mats, particularly using water as the condensing agent. In this work, we introduce a simple and scalable ethanol and water-assisted approach: by wetting electrospun PAN yarns with ethanol and water and then allowing them to dry, we harness capillary action generated during evaporation to drive the fibers into close-packed, highly aligned configurations, thereby enhancing interfiber adhesion and load transfer. This approach stands in contrast to previous methods that typically rely on organic solvents or mechanical manipulation, and represents a new, environmentally benign post-processing route for nanofiber assembly densification.

At the nanoscale, the mechanical behavior of fiber assemblies is fundamentally governed by interfiber adhesion and frictional interactions [24]. Recent work by Das and Chasiotis [9, 25] has established that PAN nanofibers exhibit pronounced rate-dependent adhesive and frictional forces approaching the shear yield strength of the polymer. Although these interactions are predominantly van der Waals-based, their magnitude is sufficient to significantly impede fiber sliding and separation under mechanical loading. Thus, understanding and optimizing interfiber contact mechanics is crucial for realizing the full mechanical potential of nanoscale fiber assemblies.

In this study, we systematically explore water-assisted capillary densification as a scalable and efficient approach for enhancing the mechanical performance of electrospun PAN nanofiber yarns. We demonstrate that controlled water exposure substantially improves yarn tensile modulus and strength through structural consolidation and enhanced interfiber adhesion. Remarkably, yarns with diameters smaller than 10 μm approach the mechanical performance of single nanofibers, indicative of highly effective stress transfer and minimized fiber slippage. Furthermore, thermal annealing after capillary treatment further elevates yarn stiffness, though with some reduction in ductility. Our findings provide direct evidence that capillary-driven structural evolution can effectively bridge the gap between the intrinsic mechanical properties of individual nanofibers and macroscopic yarn performance. This scalable post-processing technique thus opens new opportunities for the development of high-strength fiber assemblies suitable for structural and advanced functional applications.

## 2. Materials and Methods

### 2.1. Electrospinning Process and Post-treatment

Polyacrylonitrile (PAN) with an average molecular weight of ~150,000 g/mol and N,N-Dimethylformamide (DMF) were procured from Sigma-Aldrich. A 10 wt.% PAN solution was prepared by dissolving the polymer in DMF under magnetic stirring until complete dissolution. Electrospinning was carried out at room temperature under ambient conditions using a modified dual-syringe setup, as shown in Fig. 1a.

The electrospinning parameters were as follows:

- Applied voltage: 7.8 kV
- Tip-to-collector distance: 15 cm
- Syringe pump flow rate: 2 mL/hr
- Syringe angle: 45° relative to the horizontal
- Rotating funnel speed: 300 rpm
- Spool collection speed: 100 rpm (angular), with an additional linear translation of 2 mm/s to promote spatial separation of yarns.

This configuration was not optimized for producing twisted or helical yarns. Instead, the objective was to fabricate loosely assembled nanofiber bundles to assess the effect of various post-processing treatments on their mechanical properties.

Three post-processing strategies were employed to densify the electrospun PAN yarns and enhance fiber compaction. For liquid-assisted rolling, the yarn was placed on a PTFE (Teflon) tape–lined glass slide and a small droplet of solvent (deionized water or ethanol) was dispensed onto it. A second microscope slide wrapped with PTFE tape was then stroked transversely (perpendicular to the yarn axis) under light contact to distribute the solvent uniformly and wick away excess. Next, a thin PTFE-coated glass rod was rolled axially (parallel to the yarn), intermittently rotating the sample by ~90° between passes to suppress cross-sectional ovalization and maintain a near-circular profile. These low-force steps promote uniform wetting, capillarity-driven bundling/compaction, and rapid solvent drainage, yielding more reproducible outcomes than immersion-only capillary treatments. After liquid-assisted rolling, both water- and ethanol-treated yarns were placed in a vacuum desiccator for 48 hours to ensure complete removal of residual solvent and moisture. The heat-treated group received a subsequent thermal consolidation step: water and ethanol-rolled yarns were thermally annealed at 110 °C for 5 hours to promote partial bond formation at nanofiber junctions. Control tests showed that lightly rolling dry as-spun yarns did not change their mechanical response (Fig. S1, Supplementary Information), so no separate "rolled as-spun" cohort is reported. Based on the processing history, four distinct yarn types were prepared:

1. **As-spun (or Dry-rolled) yarns:** As-spun electrospun yarns without (or with) post-processing (rolling).
2. **Water-rolled yarns:** Yarns subjected to water-assisted bundling and drying.
3. **Ethanol-rolled yarns:** Yarns treated with ethanol rolling and drying.
4. **Heat-treated yarns:** Yarns that underwent thermal annealing post liquid-rolling.

### 2.2. Mechanical Testing

Tensile tests were performed using a custom-built micro-tensile setup (Fig. 1b), equipped with Newport piezoelectric actuators for displacement control and a Futek LPM200 10g load cell for force measurement. Yarn specimens were cut to a gauge length of 10 mm and mounted using two-part epoxy adhesive to prevent slippage during loading. Prior to testing, all samples were conditioned at $(25 \pm 2)$ °C and 65% relative humidity for 24 hours. The experimental setup allowed for the accurate application of tensile force while minimizing misalignment and edge-loading artifacts. The tensile tests were carried out at a nominal strain rate of $10^{-3}$ /s.

### 2.3. Image Analysis and Quantification

Morphological characterization of the PAN yarns was carried out using scanning electron microscopy (SEM). Quantitative image analysis was performed in ImageJ, employing two complementary methods: (i) Fourier component analysis and (ii) orientation analysis using the OrientationJ plugin [26]. Together, these approaches allowed assessment of fiber alignment, orientation distribution, anisotropy, and packing density, thereby enabling a direct correlation between microstructure and mechanical performance.

#### 2.3.1. Fourier Component Analysis

Structural anisotropy of the yarns was examined through Fourier spectrum analysis of SEM images. In this approach, oriented fiber assemblies generate anisotropic features in the Fourier power spectrum, with the dominant intensity distribution rotated by 90° relative to the fiber orientation in the original image. Thus, for example, horizontally aligned fibers yield a vertically elongated Fourier spectrum. The magnitude of the Fourier transform quantifies the relative contribution of different spatial frequencies and orientations, and this distribution is analyzed to extract alignment metrics. Although the Fourier transform also contains phase information, orientation analysis relies primarily on the power spectrum. To improve robustness, the ImageJ plugin subdivides each SEM image into smaller square regions and computes local Fourier spectra, thereby enabling both global and spatially resolved characterization of fiber alignment and periodicity.

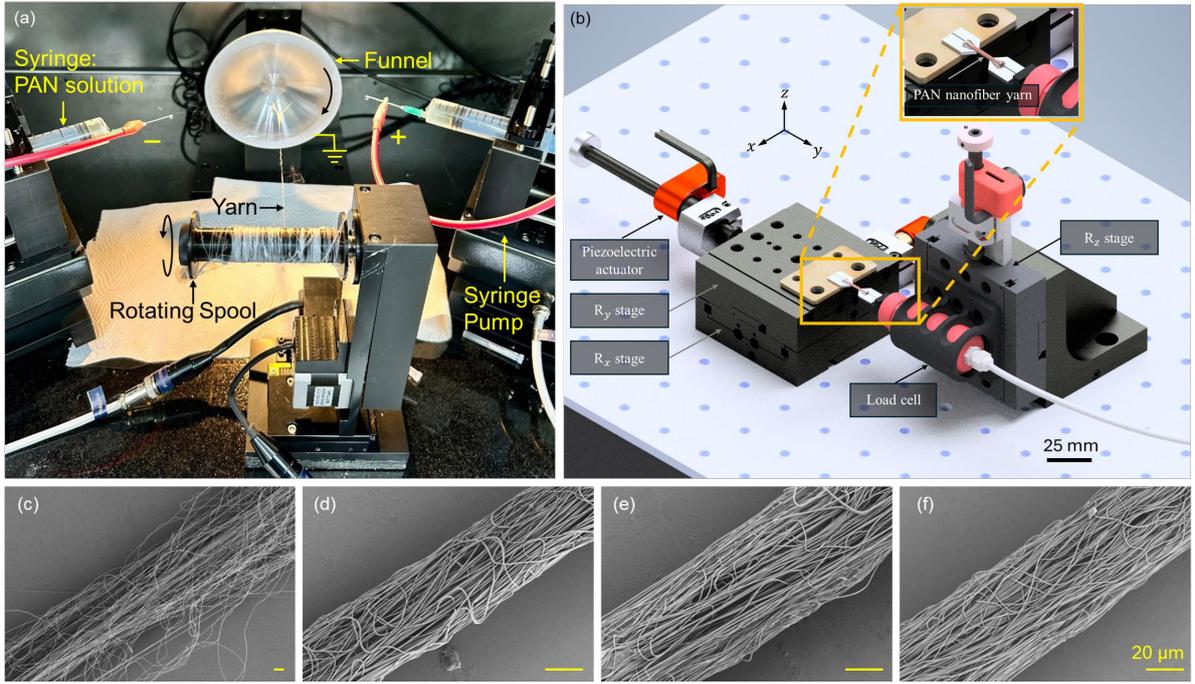

**Figure 1**. **(a)** Electrospinning apparatus with various parts labelled. **(b)** Schematic of the custom micro-tensile setup with a 10 g-capacity load cell and piezoelectric actuators for precise displacement control. (c–f) Representative SEM micrographs of yarns after **(c)** as-spun, **(d)** water-rolled, **(e)** ethanol-rolled, and **(f)** water-rolled + heat-treated processing. Scale bars: 20 µm (all SEM panels).

### 2.3.2. OrientationJ Analysis

Fiber alignment and orientation distribution were quantified using the OrientationJ plugin in ImageJ, which implements a structure tensor approach. For an image with intensity I(x,y), the local gradients, $f_x = \partial I/\partial x, f_y = \partial I/\partial y$, are computed and used to construct the structure tensor:

$$J = \begin{bmatrix} \langle f_x, f_x \rangle_w & \langle f_x, f_y \rangle_w \\ \langle f_x, f_y \rangle_w & \langle f_y, f_y \rangle_w \end{bmatrix} \quad (1)$$

where $\langle f, g \rangle_w$ denotes a Gaussian-weighted inner product over the region of interest (ROI). Solving the eigenvalue problem $Jv = \lambda v$ yields two eigenvalues, $\lambda_{max}$ and $\lambda_{min}$, with corresponding eigenvectors. The eigenvector associated with $\lambda_{max}$ defines the dominant fiber orientation, while the orthogonal eigenvector corresponds to the direction of least variation. The explicit orientation angle can also be expressed as

$$\theta = \frac{1}{2}\tan^{-1}\left(\frac{2\langle f_x, f_y \rangle_w}{\langle f_y, f_y \rangle_w - \langle f_x, f_x \rangle_w}\right) \quad (2)$$

The degree of anisotropy was quantified using the coherency factor

$$C = \frac{\lambda_{max} - \lambda_{min}}{\lambda_{max} + \lambda_{min}} \quad (3)$$

with $C \to 0$ corresponding to isotropic regions and $C \to 1$ reflecting strongly aligned structures. In addition, the gradient energy was defined as $E = trace(J) = \lambda_{max} + \lambda_{min}$, providing a measure of the local contrast. This framework enables a rigorous, reproducible quantification of orientation, anisotropy, and contrast in PAN yarns, allowing direct comparison of alignment across different treatment conditions.

### 2.3.3. Density Analysis

Fiber packing density was quantified from plan-view SEM micrographs by binary image analysis. Each image was flat-field–corrected, lightly smoothed, and segmented using adaptive local thresholding (window size fixed across images; visually validated) to mitigate low-frequency shading/charging gradients, with window size held constant across images. For each yarn, we analyzed ≥ 5 ROIs of fixed physical size (100 μm × 100 μm) at constant magnification and imaging conditions. Packing density was computed as the fraction of white (fiber) pixels within the ROI; values were averaged per sample and per condition. This in-plane metric is sensitive to kinks, loose fibers, and local voids that influence inter-fiber contact and frictional load transfer. In our dataset, in-plane fiber packing density varied monotonically with strength and, together with alignment, explained variability across processing routes, Section 3.1. We therefore reported in-plane fiber packing density as the primary structure descriptor.

Cross-sectional imaging of the as-spun yarns was not dependable: the bundle frequently drifted under the electron beam, as loosely held surface fibers are perturbed by electron impingement, leading to motion artifacts. Consequently, we opted to quantify packing from plan-view SEM images. In principle, embedding in a polymer binder (e.g., PDMS) could suppress drift and enable cross-sectional imaging; however, binder–fiber wetting/infiltration can redistribute voids and bias the measured area fraction. For this reason, we did not attempt binder-assisted cross sections here.

### 2.4. Constitutive Modeling

Electrospun yarns are hierarchical fibrous assemblies consisting of thousands of nanofibers bound together by van der Waals forces, electrostatic interactions, and mechanical entanglements. Unlike idealized parallel fiber bundles, yarns exhibit fiber waviness, misalignment, kinking, and distributed slack, which result in a progressive recruitment of fibers under tensile loading and junction-mediated load transfer (adhesion and friction), yielding a characteristic five-regime stress–strain phenomenology: an initial geometric uncrimping, an essentially linear segment, a recruitment-dominated transition, a stress quasi-plateau, and damage-controlled softening. To capture these features, we formulate a stochastic mean-field constitutive framework that accounts for distributed recruitment, statistical failure, orientation averaging, and post-failure load retention, enabling faithful reproduction of the observed macroscopic response.

The recruitment and ultimate failure of fibers are assumed to be stochastic, reflecting variability in slack, imperfections, and strength. The recruitment stretch of fiber $i$, denoted $\lambda_s^{(i)}$, and the ultimate failure stretch, denoted $\lambda_f^{(i)}$, are drawn from Weibull distributions [27, 28]:

$$\lambda_s^{(i)} = \gamma_s + \beta_s \left[ -\ln\left(1 - G_s^{(i)}\right) \right]^{1/\alpha_s} \tag{4a}$$

$$\lambda_f^{(i)} = \gamma_f + \beta_f \left[ -\ln\left(1 - G_f^{(i)}\right) \right]^{1/\alpha_f} \tag{4b}$$

where $\alpha_s$, $\beta_s$, and $\gamma_s$ are Weibull shape, scale, and threshold parameters for recruitment, and $\alpha_f$, $\beta_f$, and $\gamma_f$ are the Weibull parameters for failure. $G_r(i)$ and $G_f(i) \sim U(0,1)$ are uniform random variables. This formulation ensures gradual fiber engagement and staggered failure, producing smooth yarn-level stress–strain curves.

Because fibers are not perfectly aligned with the yarn axis, the macroscopic axial strain $\varepsilon$ must be projected onto each fiber according to its orientation angle $\theta_i$. The effective stretch experienced by fiber is therefore defined as

$$\lambda_{eff}^{(i)} = 1 + \varepsilon \cos^2(\theta_i) \tag{5}$$

This projection reflects the fact that misaligned fibers carry reduced axial strain compared to aligned ones. Equation (5) defines the effective stretch of an individual fiber with orientation $\theta_i$. Since the full orientation distribution is not tracked, we adopt a mean-field surrogate (Eq. 6), replacing the explicit $\cos^2(\theta_i)$ term by its statistical average $\phi = \langle \cos^2 \theta \rangle$, leading to the orientation-averaged effective stretch

$$\bar{\lambda}_{eff} = 1 + \phi\varepsilon \tag{6}$$

This replacement approximates the ensemble of fiber orientations by their statistical mean, simplifying the homogenization step while retaining the essential misalignment effect.

Each fiber obeys an elastic–plastic law with post-failure softening. Experimental data demonstrate elastic–perfectly plastic behavior in PAN nanofibers, with essentially no strain hardening beyond yield [29]. The yield stretch is defined relative to recruitment as:

$$\lambda_y^{(i)} = \lambda_s^{(i)} \left(1 + \frac{\sigma_0}{E_f}\right) \tag{7}$$

where $E_f$ is the nanofiber modulus and $\sigma_0$ the plastic plateau stress. The constitutive law for fiber $i$ under $\lambda_{eff}^{(i)}$ is given by

$$\sigma^{(i)}\left(\lambda_{eff}^{(i)}\right) = \begin{cases} 0 & \lambda_{eff}^{(i)} < \lambda_s^{(i)} \\ E_f\left(\dfrac{\lambda_{eff}^{(i)}}{\lambda_s^{(i)}} - 1\right), & \lambda_s^{(i)} \leq \lambda_{eff}^{(i)} < \lambda_y^{(i)} \\ \sigma_0, & \lambda_y^{(i)} \leq \lambda_{eff}^{(i)} < \lambda_f^{(i)} \\ \eta\sigma_{\text{pf}}^{(i)} \exp\left(-\dfrac{\lambda_{eff}^{(i)} - \lambda_f^{(i)}}{\delta}\right), & \lambda_{eff}^{(i)} \geq \lambda_f^{(i)} \end{cases} \quad (8)$$

Here, $\eta \in [0,1]$ controls the magnitude of post-failure load retention, and $\delta > 0$ defines the decay length. $\eta$ represents the residual bridging or frictional resistance provided by partially failed fibers, while $\delta$ sets the characteristic strain over which this contribution decays. Continuity at the failure point is maintained by defining the post-failure stress scale as

$$\sigma_{\text{pf}}^{(i)} = \min\left\{E_f\left(\frac{\lambda_f^{(i)}}{\lambda_s^{(i)}} - 1\right), \sigma_0\right\} \quad (9)$$

ensuring a smooth transition between plastic plateau and softening regimes.

The macroscopic stress is obtained through ensemble averaging over all fibers, scaled by a packing and load-transfer efficiency factor $\psi \in [0,1]$:

$$\sigma_{\text{yarn}}(\varepsilon) = \psi \frac{1}{N} \sum_{i=1}^{N} \sigma^{(i)}\left(\lambda_{eff}^{(i)}\right) \quad (10)$$

Equivalently, under the orientation-averaged approximation, this becomes

$$\sigma_{\text{yarn}}(\varepsilon) \approx \psi \langle \sigma^{(i)}\left(\bar{\lambda}_{\text{eff}}\right) \rangle \quad (11)$$

where the angular brackets denote ensemble averaging over the Weibull-distributed recruitment and failure stretches. In this manner, the model explicitly distinguishes per-fiber effective stretch from its orientation-averaged surrogate, thereby preserving physical clarity. By combining stochastic recruitment and failure, orientation reduction, efficiency scaling, and post-failure softening, the model captures the hallmark nonlinear stress–strain features of electrospun yarns. This provides a predictive and mechanistically grounded framework suitable for comparison with experiments.

## 3. Results and Discussion

### 3.1. Mechanical Properties of Treated & Untreated Yarns

The measured strain rate from all tensile tests was $1.5\times10^{-3} \pm 2\times10^{-4}$ /s. Fig. 2 shows that the stress–strain responses of as-spun and liquid-treated nanofiber yarns can be broadly divided into five distinct mechanical regimes:

Regime I – Toe–heel region: The initial low-stiffness response arises from the straightening of kinks and waviness in the yarn. Deformation is primarily geometric at this stage, with little actual stretching of the nanofibers.

Regime II – Linear elastic regime: Once the initially straightened fibers are fully aligned, the yarn responds elastically. These engaged fibers share load through axial stretching, stabilized by adhesive and frictional contacts. However, not all fibers are engaged yet; some remain slack or misoriented.

Regime III – Non-linear transition regime: At higher strains, the stress–strain curve departs from linearity due to the combined effects of progressive fiber recruitment and the onset of inelastic processes. Additional fibers begin to engage, but this is accompanied by localized sliding at fiber–fiber junctions and plastic deformation of already-stretched fibers. The interplay of stiffening (from recruitment) and softening (from inelasticity) produces the non-linear response.

Regime IV – Extended plateau regime: Over a broad strain interval, the yarn exhibits a nearly constant stress. This plateau reflects a balance between competing mechanisms: new fiber recruitment and adhesive load transfer maintain macroscopic stress, while sliding, plastic deformation, and early-stage fiber failure and disengagement/pull-out act to reduce it. Because these processes dynamically counteract one another, the yarn sustains large strains at an almost constant stress level.

Regime V – Progressive failure regime: At very large strains, recruitment is almost exhausted, and disengagement/rupture of fibers dominates. Stress decreases progressively as subsets of fibers fail in sequence. The yarn undergoes extensive thinning, often retaining structural continuity through residual adhesive bridges, but with negligible load-bearing capacity.

The principal distinction between the mechanical responses of as-spun and liquid-treated PAN yarns lies in their stress-bearing capability and failure strain. As-spun yarns exhibit lower modulus, reduced yield and peak stresses, and substantially higher strains to failure compared with treated yarns (Table 1). In some instances, the as-spun yarns can sustain elongations approaching 200% without complete rupture. Even under such extreme deformation, the yarns undergo significant thinning but remain partially intact because adhesive interactions between individual nanofibers prevent total separation. In contrast, both water- and ethanol-treated yarns display higher stresses but reduced strains to failure, with fracture strains typically limited to ~30–50%. This transition indicates that solvent treatment, followed by evaporation, brings nanofibers into closer contact, thereby strengthening adhesive interactions at junctions. The increased intimacy of these contacts restricts the capacity for fiber sliding and rearrangement, leading to a stiffer but more brittle mechanical response.

Interestingly, water-treated yarns exhibit higher average stresses than ethanol-treated counterparts. This disparity may originate from the distinct evaporation kinetics and fiber–solvent interactions. Slow evaporation of water enables more effective capillary-driven compaction of the nanofiber network, thereby improving load transfer between fibers. Ethanol, by contrast, evaporates rapidly, limiting compaction and leaving the yarn less consolidated. Additionally, ethanol induces nanofiber swelling [30], which degrades the intrinsic fiber mechanical properties and diminishes the overall load-carrying capacity.

The same conclusion can be reached from a simple mathematical analysis. Elastocapillary compaction of the yarn reflects a balance between capillary traction and filament bending. For a filament of radius $r$ and modulus $E_f$, the bending stiffness is $B = E_f I \approx E_f(\pi r^4/4)$. When two wetted filaments approach, the capillary work of adhesion per unit area is $W = \gamma(1 + \cos\theta)$, where $\gamma$ is the liquid-air surface tension and $\theta$ the static contact angle. Using this energy measure, the characteristic elastocapillary length:

$$L_{ec} = \sqrt{\frac{B}{\gamma(1 + \cos\theta)}} \tag{12}$$

and compaction is favored when the wetted segment length $L$ satisfies the criterion, $L \gtrsim L_{ec}$, or equivalently when the elastocapillary number

$$\text{Ec} \equiv \frac{\gamma(1 + \cos\theta)L^2}{B} \gtrsim 1 \tag{13}$$

Using literature values at 25 °C, water has $\gamma_w \approx 72$ mN.m$^{-1}$ [31] whereas ethanol has $\gamma_e \approx 22$ mN.m$^{-1}$ [32]. Thus, even if ethanol nearly fully wets PAN (contact angle θ≈0°), its capillary driving term $\gamma_e(1 + \cos\theta_e) \approx 44$ mN.m$^{-1}$ remains lower than water's $\gamma_w(1 + \cos\theta_w) \approx 108$ mN.m$^{-1}$ for representative water contact angle of ~60° (morphology-dependent) [33, 34] on PAN membranes. Consequently,

$$\frac{L_{ec,w}}{L_{ec,e}} = \sqrt{\frac{\gamma_e(1 + \cos\theta_e)}{\gamma_w(1 + \cos\theta_w)}} \approx 0.64 \tag{14}$$

So, $L_{ec,w} < L_{ec,e}$, and water should drive stronger compaction (smaller $L_{ec}$, larger Ec) than ethanol for the same fiber stiffness and available wetted length. Note that, ethanol's near-zero contact angle on PAN is expected from Zisman's criterion [32]: PAN's critical surface tension is $\gamma_c \approx 50$ mN.m$^{-1}$[9, 35], so liquids with $\gamma < \gamma_c$ (e.g., $\gamma_e \sim 22$ mN.m$^{-1}$) spread on clean PAN surfaces, whereas water ($\gamma_w > \gamma_c$) shows a finite angle.

Equivalently, the capillary pressure driving consolidation scales as $\Delta P \approx 2\gamma(1 + \cos\theta)/r$, so the effective compaction—and thus stiffness/strength gains—intensifies as local radius $r$ decreases, as we will see in Fig. 3.

Getting back to the mechanical response of PAN yarns, the brittle response of these yarns is further accentuated following thermal treatment. Most heat-treated yarns exhibit only a short yielding regime (Regime III) before undergoing abrupt fracture, reminiscent of brittle failure in glassy polymers. The failure strain is markedly reduced to ~15%, although tensile strength is significantly enhanced relative to untreated yarns. The increase in tensile strength can be rationalized by the thermal history of the yarns. Annealing at 110 °C, a temperature close to the glass transition of PAN ($T_g \approx 85$–$105$ °C), promotes localized fusion of adjacent nanofibers at their contact points. This thermal consolidation densifies the fibrous network, and further restricts fiber mobility. The resulting microstructural changes enhance stiffness and load transfer efficiency, thereby increasing the overall strength of the yarn. Post-mortem SEM analysis confirmed the presence of fused junctions between neighboring fibers, corroborating this interpretation. However, such fusion also suppresses energy-dissipating mechanisms—specifically inter-fiber sliding and progressive fiber disengagement—that are essential for sustaining large deformations. Consequently, while strength is elevated, the loss of these ductility-governing mechanisms accounts for the abrupt and brittle fracture observed in heat-treated yarns.

Figure 2(f) presents the median stresses together with the median absolute deviation obtained from multiple tensile tests on the various yarns. Plotting the median rather than the mean reduces sensitivity to outliers and provides a more representative measure of the central response for these highly variable fibrous assemblies. The consolidated comparison clearly highlights the effect of liquid treatment: water treatment yields the highest strengths while still retaining appreciable ductility, in contrast to ethanol-treated or heat-treated yarns.

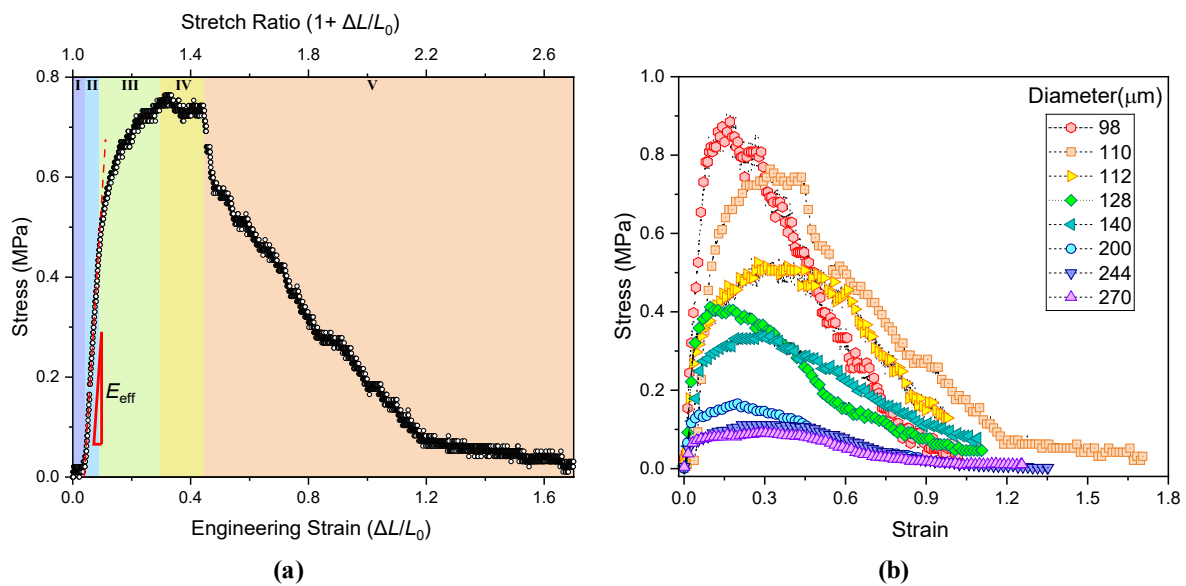

(a)   (b)

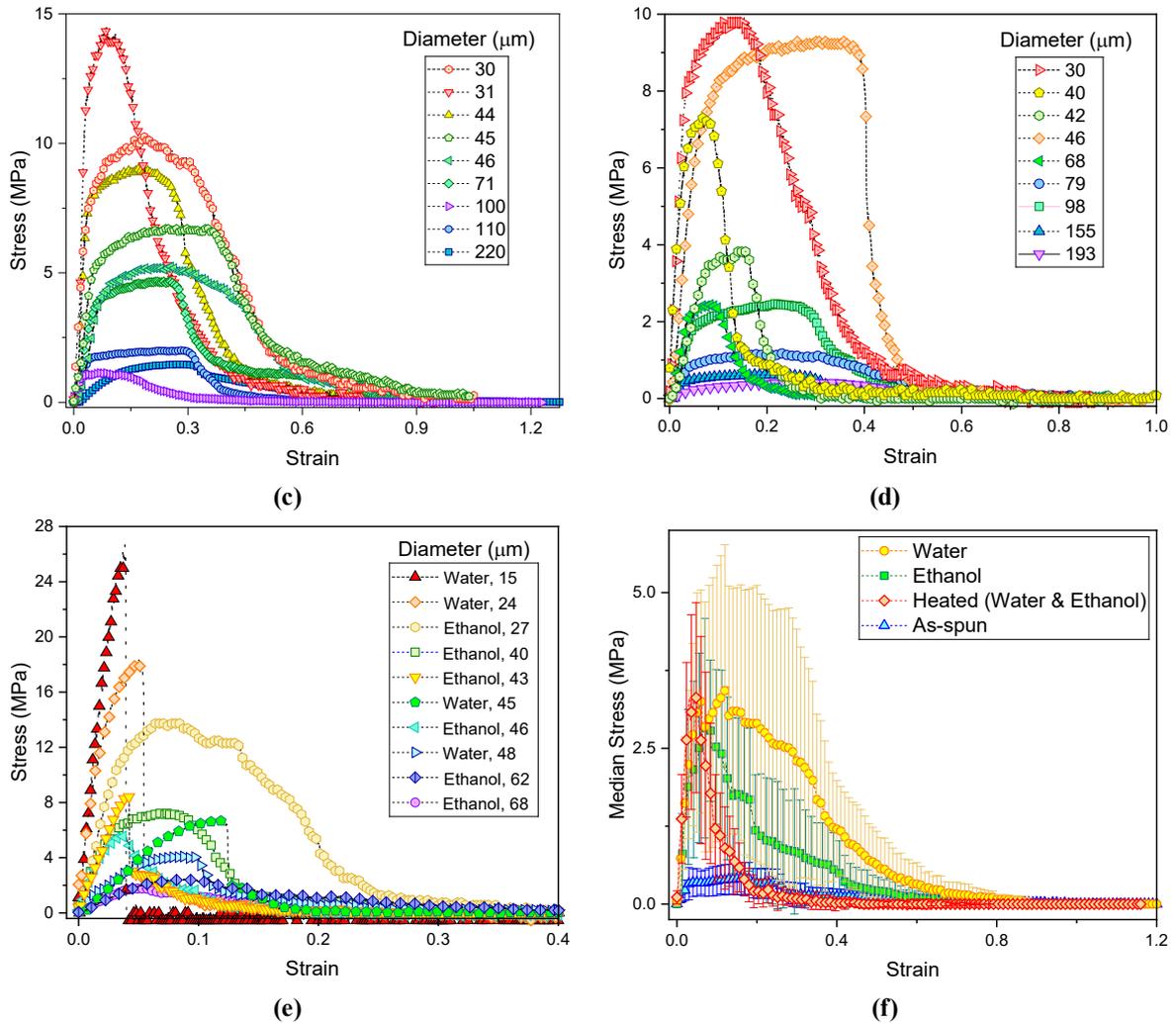

**Figure 2.** (a) Representative stress–strain curve for an as-spun yarn, highlighting five regimes: I—toe-heel; II—linear elastic; III—non-linear transition; IV—extended plateau; V—progressive failure. (b–e) Individual stress–strain responses for (b) as-spun, (c) water-rolled, (d) ethanol-rolled, and (e) liquid-rolled + heat-treated yarns, illustrating the processing-dependent changes in stiffness, peak stress, and post-peak behavior. (f) Summary statistics across replicates: median stresses with error bars showing the median absolute deviation (MAD). The strain rate used in the study was ~ $10^{-3}$ /s.

**Table 1.** Across-tests summary: average mechanical properties of nanofiber yarns

| Type (# Samples) | Peak Load (mN) | | Peak Stress (MPa) | | (σ-ε) Area (MJ/m³) | | $E_{eff}$ (MPa) | |
|---|---|---|---|---|---|---|---|---|
| | Mean | CV(%) | Mean | CV(%) | Mean | CV(%) | Mean | CV(%) |
| As-spun (N=18) | 7.3 | 72.6 | 2.1 | 207.1 | 0.6 | 152.0 | 79 | 240.8 |
| Ethanol (N=14) | 10.5 | 59.6 | 5.2 | 114.6 | 1.2 | 105.2 | 143 | 130.7 |
| Water (N=24) | 15.0 | 109.8 | 5.9 | 106.9 | 1.5 | 76.4 | 163 | 148.5 |
| Water (heated) (N=12) | 8.2 | 35.0 | 7.8 | 92.8 | 0.6 | 128.3 | 211 | 124.1 |
| Ethanol (heated) (N=7) | 8.1 | 33.3 | 6.2 | 68.4 | 0.6 | 113.6 | 211 | 60.3 |

Figure 2 also demonstrates that both the peak stress and effective modulus of the yarns increase systematically with decreasing yarn diameter. To capture this trend more clearly, the effective modulus and peak strength are plotted as a function of yarn diameter in Fig. 3(a). A striking observation is that both the effective modulus as well as the peak strength exhibit a strong inverse dependence on diameter, with modulus and peak stress ranging from ~1.1 GPa, 30 MPa for yarns of 10 μm diameter to ~3 MPa, 0.12 MPa for those of 245 μm diameter. Thus, thinner yarns are markedly more load-efficient. Indeed, for yarn diameters <10 μm, their mechanical response approaches that of single PAN nanofibers: literature reports for similar diameter nanofibers (~500-nm) tested at similar strain-rates (~$10^{-3}$ s$^{-1}$) show $E \approx 1.5$ GPa and $\sigma_{peak} \approx$ 30–60 MPa [29]. Although part of this trend can be attributed to geometric normalization by cross-sectional area—which accentuates the apparent stresses in liquid-treated yarns that shrink during solvent evaporation—this effect alone cannot fully explain the observed strengthening. As demonstrated in the following paragraphs, additional factors such as enhanced fiber compaction, improved alignment, reduced porosity, and increased inter-fiber adhesion also contribute significantly to the superior load-carrying capability of liquid-treated yarns. The observed scaling behavior, which to our knowledge has not been previously reported for PAN nanofiber yarns, underscores the critical role of mesoscale organization—fiber alignment, compaction, and contact density—in dictating their mechanical performance.

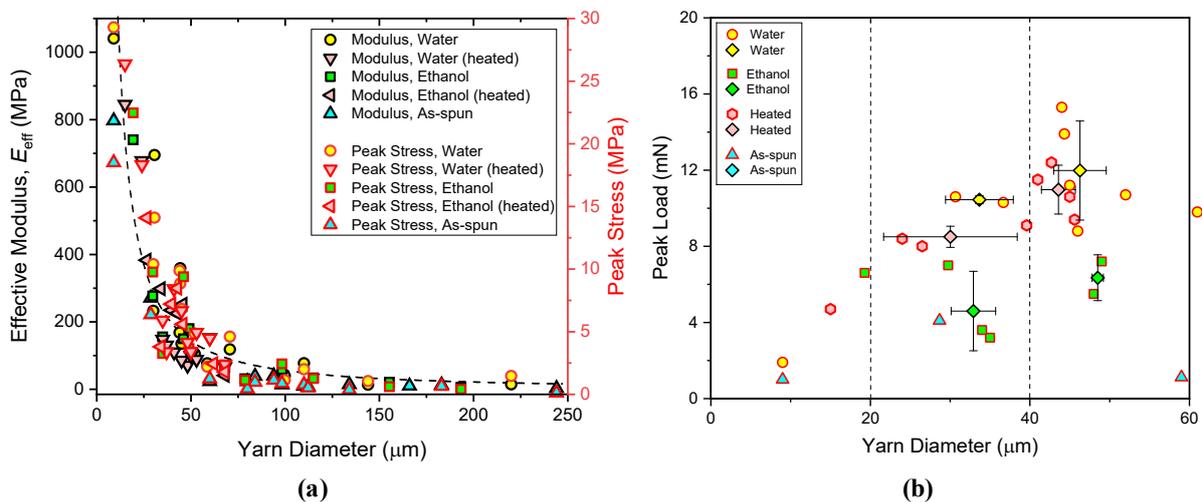

**Figure 3. (a)** Effective modulus and peak stress plotted against yarn diameter; **(b)** Peak load for the different processing routes, grouped into three diameter ranges. Across all bins, water-rolled yarns (with or without subsequent heat treatment) consistently exhibit higher load-carrying capacity than as-spun yarns.

To disentangle the influence of surface treatment from diameter-dependent effects, the yarns were first categorized into three diameter bins (0–20 μm, 20–40 μm, and 40–60 μm), which encompass the range where most of the stiffening and strengthening effects are observed. Within each bin, the peak load distributions for the different treatment groups were compared (Fig. 3(b)). Across all diameter ranges, treated yarns (water-, ethanol-, and heat-treated) consistently exhibited higher peak loads than their as-spun counterparts. It is important to emphasize that load, rather than stress, is used here because stress values can be artificially inflated by diameter shrinkage associated with solvent evaporation, whereas peak load more directly reflects the intrinsic load-bearing capacity of the yarn. The consistent enhancement in peak load across diameter bins demonstrates that treatment effects cannot be explained solely by geometric scaling; instead, treatment improves inter-fiber interactions and load transfer efficiency.

To further isolate the effect of treatment while eliminating geometric variability, a single PAN yarn was cut into four equal segments, each subjected to a different condition: as-spun, water-treated, ethanol-treated, and water + heat-treated. This controlled approach ensures identical starting geometry and fiber distribution, allowing a direct comparison of treatment effects. Portions of each segment were analyzed using SEM to assess microstructural changes such as fiber alignment, packing density, and evidence of fusion, while the remaining segments underwent tensile testing to evaluate mechanical response.

SEM micrographs (Fig. 4(a–d)) and corresponding orientation color maps (Fig. 4(e–h)) reveal a clear enhancement in fiber alignment and packing density upon liquid rolling and subsequent thermal annealing. Quantitative image analysis, carried out using two complementary approaches—2D Fourier transforms and OrientationJ-based orientation mapping—confirmed an increase in anisotropy and directional coherence in the treated yarns. These microstructural changes directly correlate with the observed improvements in mechanical performance. Directionality histograms derived from OrientationJ (Fig. 4(i–l)) reveal pronounced differences among treatments. The untreated yarn displayed sparse, disordered fibers with poor inter-fiber contact, consistent with limited load-sharing capability. By contrast, liquid- and heat-treated yarns exhibited more compact structures with markedly improved alignment and reduced orientation spread, features that show enhance inter-fiber adhesion and facilitate efficient load transfer during deformation.

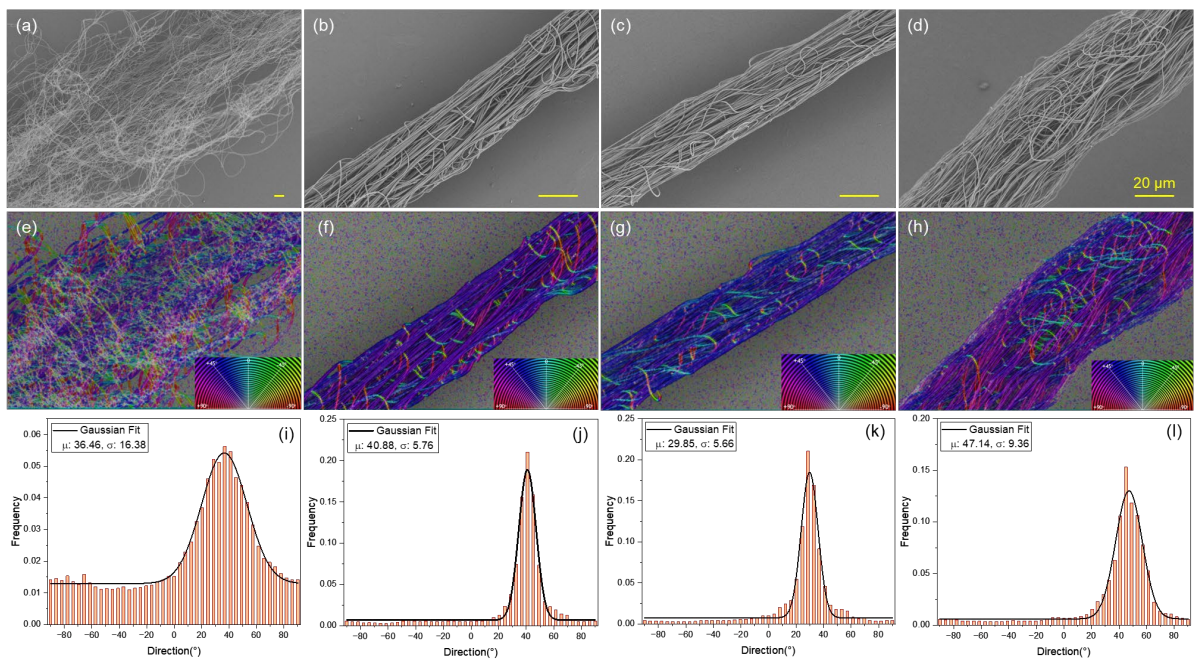

**Figure 4.** SEM micrographs from adjacent segments of the same parent yarn after **(a)** as-spun, **(b)** ethanol-rolled, **(c)** water-rolled, and **(d)** water-rolled + heat-treated processing. Scale bars: 20 µm. **(e–h)** OrientationJ directionality maps for (a–d): the rainbow hue encodes local fiber orientation. Treatment improves the fiber alignment along the yarn axis. **(i–l)** Corresponding polar orientation histograms (Fourier-based angular distributions) for (a–d), showing sharpening about the principal (yarn) axis with processing.

The load–displacement curves (Fig. 5(a)) highlight the strong influence of surface treatment on yarn mechanics. The water-treated yarn (51 μm diameter) carried the highest peak load, closely followed by the water + heat-treated yarn (50 μm). By contrast, the dry as-spun yarn (134 μm) exhibited the lowest load-bearing capacity, reflecting its sparse, weakly compacted microstructure with limited inter-fiber adhesion. The ethanol-treated yarn (48 μm) showed intermediate performance. This behavior may arise from a competition between two opposing effects: improved fiber alignment during solvent treatment, which enhances load transfer, and nanofiber swelling in ethanol [30], which compromises intrinsic fiber stiffness and reduces load-carrying ability. As a result, ethanol treatment yields better alignment compared to the as-spun yarn but fails to translate this fully into mechanical strength. When normalized by cross-sectional area, the stress–strain curves preserve the same qualitative ranking. However, because solvent treatment markedly reduces yarn diameter, the calculated stresses for treated yarns appear significantly magnified relative to the as-spun sample. Thus, while stress normalization can obscure absolute load-bearing capability in cases where cross-sectional changes are substantial, direct comparison of load–displacement behavior provides a more reliable measure of the intrinsic strengthening imparted by surface treatments.

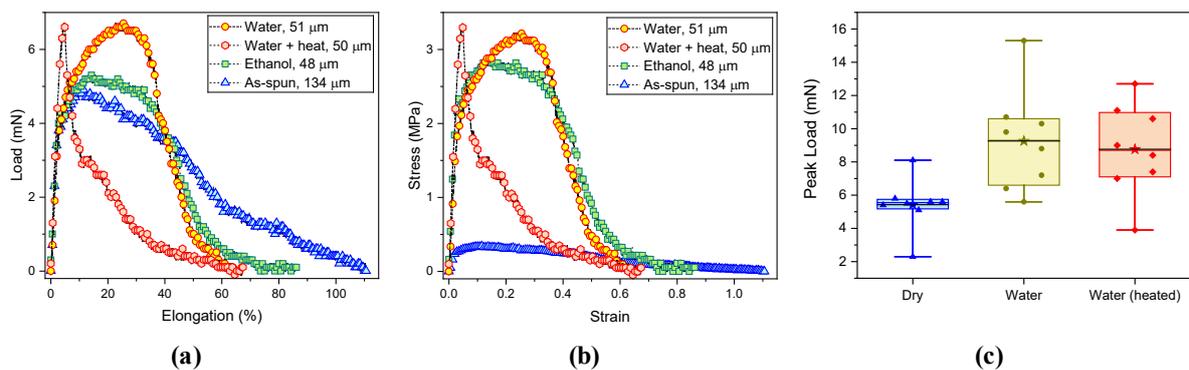

**Figure 5.** **(a)** Load–elongation response for four segments with various treatments cut out of a single yarn specimen. **(b)** Corresponding engineering stress–strain response derived from (a). **(c)** One-way ANOVA on peak load across processing conditions—as-spun, water-treated, and water-treated + heat-treated—(N = 8) shows a significant processing effect (p = 0.013), indicating treated yarns carry higher peak loads than as-spun.

Statistical analysis further supports these observations. The box plots of peak load (Fig. 5(c)) and Table 2 clearly show the strong influence of surface treatment on yarn mechanics. As-spun yarns exhibited the lowest peak loads, clustered around ~5–6 mN, while both water- and water + heat-treated yarns carried substantially higher loads with medians near 9–10 mN. On average, this corresponds to a ~70% increase relative to untreated yarns. Statistical analysis confirmed these observations: a one-way ANOVA revealed a significant overall treatment effect ($F(2,21) = 5.36, p = 0.013$), with post-hoc Tukey comparisons showing that both water-treated and water + heat-treated yarns had significantly higher peak loads than the as-spun yarns ($p < 0.05$). By contrast, the difference between water and water + heat-treated groups was not statistically significant ($p \approx 0.92$), consistent with the overlapping distributions in the box plot. Thus, while heating after water treatment does not further increase peak load, it appears to stabilize the compacted structure, as indicated by the narrower spread of data points. Overall, treatment explained ~34% of the variance in peak load, underscoring the robust impact of solvent-induced fiber consolidation on strengthening PAN yarns, even in the presence of microstructural heterogeneity.

Interestingly, the strengthening effect of water treatment in electrospun PAN yarns can be understood as a two-step process: first, the formation of liquid menisci generates capillary forces that actively draw neighboring nanofibers into closer contact; second, as water evaporates, these menisci shrink, further amplifying the capillary pull and leaving the fibers tightly bundled. This densification enhances inter-fiber adhesion and alignment, thereby improving load-sharing efficiency across the yarn. A parallel can be drawn to traditional silk processing, where natural fibers are softened in water and subsequently combined into threads. Although the underlying molecular mechanisms differ—capillary-driven compaction and adhesion in PAN versus sericin-mediated binding in silk—the common outcome is that water facilitates intimate filament contact and enhances collective mechanical performance.

**Table 2.** Average mechanical properties of treated and untreated segments (cut out from same yarn)

| Type (# Samples) | Peak Load (mN) | | Peak Stress (MPa) | | ($\sigma$-$\varepsilon$) Area (MJ/m$^3$) | | $E_{\text{eff}}$ (MPa) | |
|---|---|---|---|---|---|---|---|---|
| | Mean | CV(%) | Mean | CV(%) | Mean | CV(%) | Mean | CV(%) |
| Unrolled (N=8) | 5.4 | 28.9 | 0.5 | 86.0 | 0.1 | 57.7 | 13 | 96.1 |
| Ethanol (N=3) | 7.2 | 23.0 | 3.0 | 22.0 | 0.9 | 62.6 | 118 | 49.3 |
| Water (N=8) | 9.3 | 33.1 | 5.7 | 48.7 | 1.6 | 49.6 | 94 | 48.2 |
| Water (heated) (N=8) | 8.8 | 33.7 | 4.2 | 32.6 | 0.4 | 52.6 | 95 | 36.4 |

To establish a quantitative link between microstructure and mechanical response, correlation analyses were performed between peak load and two key SEM-derived metrics: fiber alignment (coherency factor) and in-plane fiber packing density (Fig. 6(a–b)). Both parameters exhibited strong positive correlations with peak load, with $R^2=0.68$ ($p = 0.001$) for alignment and $R^2=0.75$ ($p < 0.001$) for packing density. These results demonstrate that improved orientation and compaction of nanofibers are both critical to enhancing the load-bearing efficiency of PAN yarns.

Representative examples (Fig. 6) illustrate the strength of these correlations. Sample 1, with a relatively low alignment (coherency factor) of 38%, sustained only 6 mN of load. Following water treatment, alignment improved to 75%, accompanied by an 84% increase in peak load, 11 mN. In another case, Sample 3, which was already moderately aligned (60%), showed a further 37% increase in load-bearing capacity, from 8 mN–11 mN, after water + thermal treatment, attributable primarily to densification and partial fiber fusion. These comparisons underscore that both alignment and compaction act synergistically: alignment reduces stress concentrations by ensuring more uniform strain distribution across fibers, while higher density enhances inter-fiber load transfer through increased frictional and adhesive interactions (van der Waals forces, and mechanical interlocking). The reduction of internal voids further facilitates stress bridging, thereby improving macroscopic strength.

Taken together, these findings confirm that surface treatments significantly enhance the mechanical properties of PAN yarns by simultaneously improving fiber alignment and packing density. More broadly, they establish a direct structure–property relationship, demonstrating that yarn strength scales with microstructural anisotropy and compaction. Such quantitative correlations represent an important step toward a micromechanical understanding of electrospun PAN yarns, providing a framework for rational design and optimization of nanofiber-based structural materials.

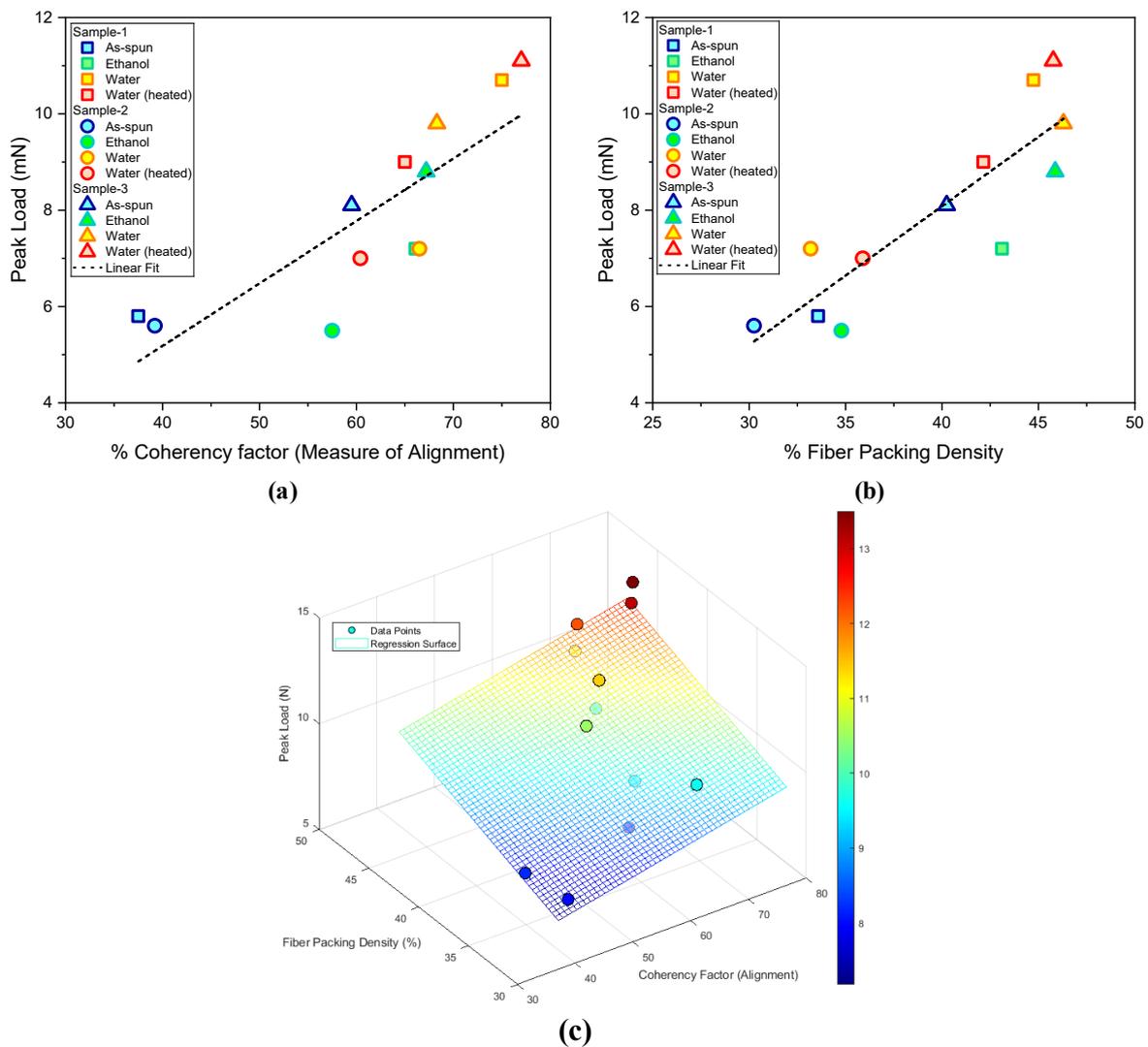

**Figure 6. (a)** Peak load versus image-based coherency factor (OrientationJ), a proxy for in-plane fiber alignment. **(b)** Peak load versus planar fiber packing density (SEM area fraction). Points are individual specimens; lines indicate least-squares fits. **(c)** Joint model of peak load as a function of coherency and planar density; the surface shows the multiple-regression fit with data points overlaid. Both alignment and density are strong individual predictors of peak load, and their combination further increases explanatory power, indicating complementary contributions.

To further assess the combined influence of fiber alignment and packing density on yarn mechanics, a multiple linear regression model was employed:

$$P = \beta_0 + \beta_1 A + \beta_2 D + \delta \tag{15}$$

where $P$ is the peak load (mN), $A$ is the fiber alignment quantified by the coherency factor (0–1), $D$ is the fiber density expressed as the fiber area fraction, $\beta_0$ is the intercept, $\beta_1$ and $\beta_2$ are the regression coefficients for alignment and density respectively, and $\delta$ represents the residual error.

This framework enables decomposition of the variance in peak load into contributions from alignment and density, thereby distinguishing their independent effects. The overall explanatory power of the model was quantified using the coefficient of determination:

$$R^2 = 1 - \frac{\sum (P_i - \hat{P}_i)^2}{\sum (P_i - \bar{P})^2} \tag{16}$$

where $P_i$ is the measured peak load, $\hat{P}_i$ the model-predicted value, and $\bar{P}$ the mean. Partial regression coefficients ($\beta_1$, $\beta_2$) indicate the relative weight of alignment versus density in predicting yarn strength. Statistical significance of each predictor was evaluated using standard $F$-tests and associated $p$-values.

Interpretation of the regression results focuses on three aspects: (i) whether both alignment and density emerge as significant independent predictors of strength, (ii) the magnitude of their coefficients relative to each other, and (iii) whether the combined model achieves a higher $R^2$ compared to individual bivariate correlations (alignment vs. load; density vs. load). A significant improvement in explanatory power would imply that alignment and density act as complementary descriptors of yarn microstructure, each capturing distinct features of fiber organization that contribute to mechanical performance.

The results of a multiple regression analysis are provided in Table 3. Individually, both alignment ($R^2=0.68$, adj. $R^2=0.65$, p=0.00098) and density ($R^2=0.75$, adj. $R^2=0.72$, p=0.00028) emerged as strong predictors of peak load, with density exerting a slightly stronger effect. When the two descriptors were combined, the explanatory power of the model increased further ($R^2=0.80$, adj. $R^2=0.76$, p=0.00066), indicating that alignment and density provide complementary rather than redundant information. The improvement in adjusted $R^2$ confirms that fiber alignment explains additional variance in strength that is not captured by density alone. These findings suggest that while fiber packing density governs load transfer efficiency through enhanced inter-fiber contact, alignment contributes independently by reducing stress concentrations and distributing strain more uniformly across the fiber network. Thus, both parameters act synergistically to control yarn strength, establishing them as coupled but non-redundant microstructural descriptors of mechanical performance.

Table 3. Results of linear regression analyses between peak load and microstructural descriptors

|  | Effect of alignment | Effect of density | Effect of alignment & density |
|---|---|---|---|
| **R-square** | 0.68 | 0.75 | 0.80 |
| **Adjusted R-square** | 0.65 | 0.72 | 0.76 |
| **F-statistic** | 21.1 | 29.9 | 18.4 |
| **p-value** | 9.82×10$^{-4}$ | 2.75×10$^{-4}$ | 6.63×10$^{-4}$ |

### 3.2. Modeling Results

The stochastic mean-field constitutive model provides a robust description of the nonlinear stress–strain behavior of electrospun PAN yarns. By incorporating distributed recruitment, elastic-perfectly plastic fibers, stochastic failure, orientation averaging, and post-failure load retention, the model successfully reproduces the characteristic mechanical regimes observed experimentally in Section 3.1. All PAN nanofibers are assumed to have an elastic modulus of 1.5 GPa and yield stress of 40 MPa and after yielding becomes a perfectly plastic material, as has been observed experimentally [29].

Model parameters $p$ were estimated by weighted non-linear least squares. For a specimen with measured stretch–stress pairs $\{(\lambda_i, \sigma_i^{\exp})\}_{i=1}^N$, we minimized the objective

$$J(p) = \underbrace{\sum_{i=1}^{N} w_i [\sigma^{\mathrm{mod}}(\lambda_i; p) - \sigma_i^{\exp}]^2}_{\text{data misfit}} + \underbrace{\sum_k \mu_k \, g_k(p)^2}_{\text{soft physics constraints}} + \underbrace{\alpha_{\mathrm{hold}} \|p - p_{ref}\|_2^2}_{\text{quadratic anchor}} \quad (17)$$

where the non-negative weights $w_i$ are user-set to emphasize the elastic rise and the main post-peak drop while gently de-emphasizing the far tail with a smooth taper. The second term encodes soft physics constraints: each $g_k(p)$ is a scalar measure of a model-consistency condition (e.g., monotonic fiber recruitment, bounded end-recruitment $0 \leq \phi \leq 1$, and smooth post-peak stress evolution), constructed so that $g_k(p) = 0$ when the condition is satisfied and $g_k(p) > 0$ when it is violated. The coefficients $\mu_k \geq 0$ (penalty weights) control the relative strength of these soft constraints. The final quadratic term $\alpha_{\mathrm{hold}}\|p - p_{ref}\|_2^2$ is a Tikhonov "anchor" that keeps the solution near a reference parameter set $p_{ref}$ (either the initial guess or the best fit from a previous stage) to improve stability and prevent overfitting; $\alpha_{\mathrm{hold}} \geq 0$ sets how strongly this anchoring acts. Component-wise bounds on $p$ enforce physical admissibility (e.g., non-negative moduli, admissible recruitment fractions). We minimized $J(p)$ using MATLAB's *lsqnonlin* (trust-region reflective) with tight tolerances (FunctionTolerance 10$^{-8}$, OptimalityTolerance 10$^{-6}$), supplying the stacked residual vector $[\sqrt{w_i}(\sigma^{\mathrm{mod}} - \sigma^{\exp}); \sqrt{\lambda_k} g_k(p); \sqrt{\alpha_{\mathrm{hold}}}(p - p_{ref})]$ so that the solver's $\ell_2$ norm matches the objective above.

Fitting proceeded in three stages. Stage A (data-only) minimized the misfit term to obtain a baseline; 1–3 jittered restarts were used to avoid poor local minima. Stage B (gentle constraints) introduced small $\lambda_k$ to guide the solution toward physically consistent fractions; the result was accepted only if the global mean-square error (MSE) did not increase by more than ~2%. Stage C (tighten) increased $\lambda_k$ moderately and was adopted only if it strictly improved the global MSE. Finally, an end-polish adjusted only tail-sensitive parameters while holding others near $p_{ref}$ (penalty $\alpha_{hold}$); three light, shape-aware terms were applied over the last 15–20% strain (endpoint anchor to the mean of the last data, "no up-slope" at the end, and mild curvature damping). Acceptance used *per-point* MSE deltas to avoid bias from unequal point counts: the end-window MSE had to improve by ≥5% while the global MSE could worsen by ≤0.3%. Fraction curves were computed from the final parameters using the model's internal state equations and verified to satisfy mass balance $R \approx f_{el} + f_y + f_{pf}$ and $0 \leq f \leq 1$ pointwise.

Figure 7(a) compares model fits with representative experimental curves. The agreement is notable across the full strain range for all the yarns whether as-spun or treated, with the model capturing all the toe-heel, elastic, nonlinear extended plateau and the absence of catastrophic rupture, consistent with the persistent fibrillar bridging seen in SEM images of failed yarns. The toe region is governed primarily by the Weibull distribution of recruitment stretches $\lambda_s$, which reflects variability in initial fiber slack and kinking. The stochastic distribution of ultimate failure stretches $\lambda_f$ produces the smooth post-peak decay, while the residual load retention factor $\eta$ accounts for incomplete separation and persistent load-bearing capacity. This is consistent with experimental observations where extreme thinning is accompanied by continued mechanical integrity.

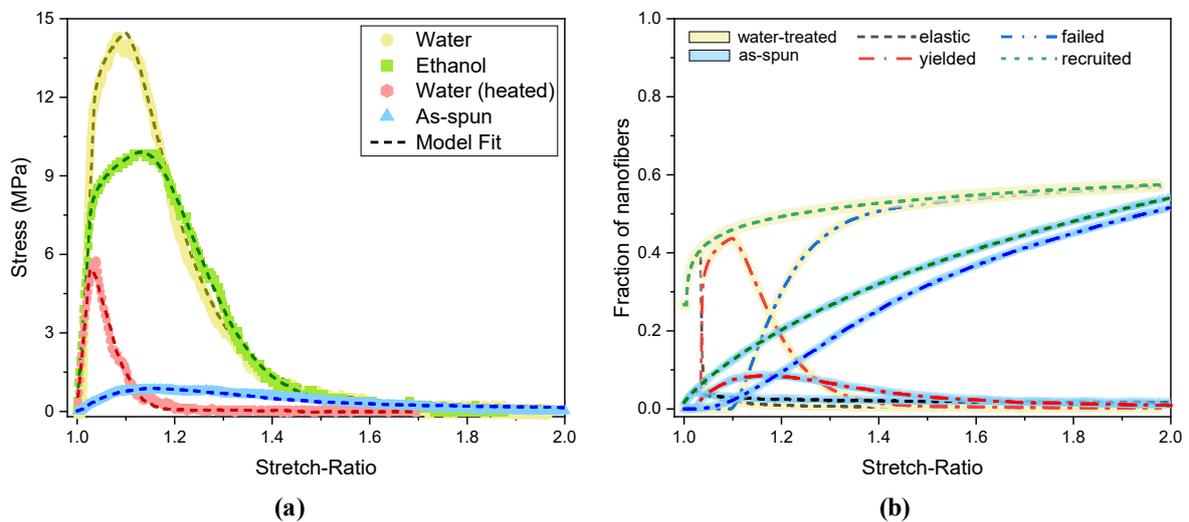

(a)    (b)

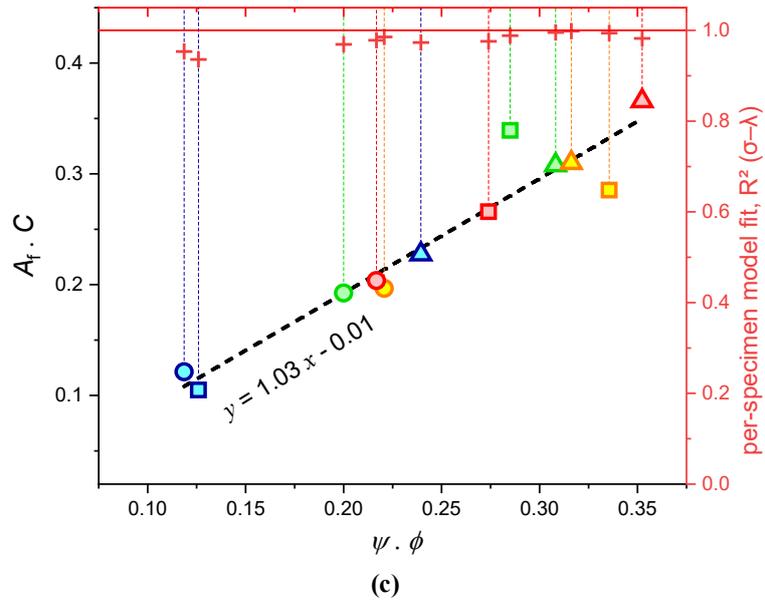

**(c)**

**Figure 7.** **(a)** Engineering stress–stretch responses ($\sigma$–$\lambda$, with $\lambda = 1 + \varepsilon$) for representative as-spun and treated yarns; dashed curves are model fits. The model faithfully captures the key deformation stages and the overall response. **(b)** Fitted fiber-state fractions versus stretch for as-spun and water-treated yarns (states: elastic, yielded, failed, recruited). **(c)** Relationship between an SEM-based microstructure proxy and model-inferred microstructure. Colored symbols: (fiber packing density (surface area fraction) × coherency) versus $\psi\cdot\phi$ from the inverse model for each specimen; dashed line is a linear fit. Red plus symbols (right axis) show the goodness of $\sigma$–$\lambda$ fits for the same specimens ($R^2\approx 1$), confirming accurate mechanical reproduction. The strong monotone trend indicates that the fitted microstructure parameter $\psi\phi$ corresponds to independently measured SEM features (packing × alignment).

Figure 7(b) shows the evolution of nanofiber state fractions for as-spun and water-treated yarns. In the water-treated case the elastic and recruited fractions are already non-zero at $\lambda=1$, indicating capillary pre-tension/compaction during treatment: liquid bridges and slow drying remove geometric slack, bring neighbors into contact, and leave a subset of filaments taut at zero load. Consequently, the recruited fraction rises earlier and more steeply (notice the slope of the recruited curve at low stretch-ratios), while the yielded fraction exhibits an earlier, narrower transient than in the as-spun yarn. Beyond the stress peak, the post-failure fraction grows in both materials, and the state fractions tend to converge at large stretch. This convergence is expected: the ultimate recruited plateau is largely constrained by the shared topology/chemistry (segment lengths, broken ends, contact network), so most segments that can bridge eventually do so in both cases. The large difference in stress nonetheless persists because water treatment increases the product of packing and alignment (considered in following paragraphs), amplifying the entire response, and it triggers earlier recruitment; together these effects produce a steeper initial rise, a higher peak, and sharper softening even though the late-stage state fractions converge. Also, note that to mitigate (but not eliminate) nonconvexity effects, we performed 1–3 jittered restarts and report the best solution; nevertheless, the fit is not guaranteed to be unique and may represent a local minimum.

Two other important parameters in the model are the orientation averaging through the parameter, $\phi=\langle\cos^2\theta\rangle$, and the packing efficiency parameter, $\psi$. $\phi$ reduces the effective stretch carried by misaligned fibers whereas $\psi$ helps capture differences in load transfer arising from changes in density. Treated yarns, which display higher alignment (coherency factor) and density in SEM and OrientationJ analyses, are reproduced by larger $\phi$ and $\psi$ values, respectively, thereby enhancing stiffness and strength. Together, $\phi$ and $\psi$ directly link the model to the structural descriptors extracted from SEM, offering a quantitative bridge between microstructure and mechanics.

Fig. 7(c) shows a strong linear relation between the SEM-derived microstructure index $A_f \cdot C$ (in-plane fiber packing density × coherency) and the model-inferred $\psi \cdot \phi$ across $n=12$ specimens that are colored by applied processing routes. A least squares fit gives slope 1.03±0.10 (SE) and intercept −0.014±0.026, with $R^2$=0.91 (adj. $R^2$=0.90), RMSE =0.024, and Pearson $r$=0.96. The slope is highly significant ($p$=1.3×10$^{-6}$), and the rank association is likewise strong (Spearman $\rho$=0.92, $p$=1.9×10$^{-5}$). Together with the near-unity goodness of the individual σ–λ fits shown on the right axis, this indicates that the inverse model's microstructure term $\psi \cdot \phi$ maps closely onto independently measured SEM features (packing × alignment). Another important observation is that the fitted trend holds within groups also. Using Hermans orientation factor, $H$ instead, yields the equivalent proxy $A_f \cdot (H+1)/2$ with slightly smaller $R^2$, consistent with the expected monotone relationship between $C$ and $H$. The small intercept difference likely reflects the surface-vs-core density mismatch and measurement scale factors.

Overall, the model not only fits the experimental data with high fidelity but also provides physical insight into the micromechanical mechanisms governing yarn strength. Variability in slack and imperfections explains the extended recruitment regime; inter-fiber adhesion and friction dominate the plateau; and partial fiber failure underlies the gradual failure tail. The framework thus establishes a predictive, mechanistically grounded tool for understanding how processing (e.g., water or thermal treatment) modifies microstructure to enhance yarn performance.

## 4. Conclusion

This work establishes a microstructure-aware pathway to strengthen electrospun PAN yarns by coupling liquid-assisted capillary densification (water or ethanol rolling) with moderate thermal annealing, and by quantitatively linking those process routes to mechanical response through imaging-derived descriptors and a stochastic, physically informed constitutive model. The post-processing routes were deliberately simple—water/ethanol rolling followed by drying and, in one branch, a 110 °C anneal of the liquid-rolled yarns—so that changes in mechanics could be traced to capillarity-driven compaction and junction evolution rather than to processing complexity. Quantitative SEM analysis provided two complementary descriptors of yarn architecture: alignment (coherency factor) and packing density (area fraction). Together they enabled a direct, image-based correlation between structure and peak load. Individually, alignment and density each showed strong explanatory power; taken together in a multiple regression, they yielded a higher adjusted $R^2$, demonstrating that the two descriptors are non-redundant and synergistic contributors to strength. This result supports a clear mechanistic picture: tighter packing enhances load transfer via increased inter-fiber contact, while improved alignment mitigates stress concentrations and distributes strain more uniformly across the network. To interpret the full stress–strain curves, we developed a stochastic mean-field model that explicitly encodes distributed fiber recruitment and failure (Weibull), elastic-perfectly plastic fiber behavior with post-failure load retention ($\eta$, $\delta$), and orientation averaging through $\phi = \langle \cos^2\theta \rangle$, with an efficiency factor $\psi$ capturing packing- and contact-mediated load transfer. The model parameters were estimated via weighted non-linear least squares under soft physics-based regularization, and the framework reproduced the key nonlinear regimes observed experimentally. Importantly, the mapping of $\phi$ and $\psi$ to image-derived alignment and density closes the structure→mechanics loop, turning micrographs into predictive inputs.

Overall, the study delivers three actionable outcomes. First, it identifies simple, scalable post-processes that tune yarn mechanics via capillary compaction and modest annealing. Second, it demonstrates that alignment ($\phi$) and packing ($\psi$) are complementary, quantifiable levers that jointly control strength. Third, it provides a calibrated constitutive tool that captures recruitment, yielding, softening, and residual bridging, enabling forward predictions under new processing conditions. The framework can be generalized to other electrospun polymers and hierarchical preforms, providing a process-to-performance map useful for yarn-reinforced composites and flexible textile applications.

## 5. Acknowledgement

The authors acknowledge support from the Anusandhan National Research Foundation (ANRF) through core research grant No. CRG/2023/006357. The authors also thank Mr. Soumyadeep Dey and Mr. Sumanth Bommali for helping with preliminary experiments and analysis during their summer internships.

## 6. Data and Code Availability

The data that support the findings of this study are available from the corresponding author upon reasonable request. Analysis scripts will be shared for academic use on request.

## 7. Conflict of Interest Statement

The authors declare no competing financial or non-financial interests.